\documentclass[numberedappendix, twocolumn]{emulateapj} 
\usepackage{rotating} 
\usepackage{amsmath,amssymb}
\usepackage[usenames,dvipsnames]{pstricks}
\usepackage{epsfig}
\usepackage{subfigure}
\usepackage{pst-grad} % For gradients
\usepackage{pst-plot} % For axes
\usepackage{color}
\usepackage{float}
\usepackage{natbib} 
\graphicspath{{figures/}}

\begin{document}

\title{Temporal Evolution of the Scattering Polarization of the Ca {\sc ii} IR triplet in \\ Hydrodynamical Models of the Solar Chromosphere.}

\author{E.S. Carlin\altaffilmark{1}, A. Asensio Ramos\altaffilmark{1,2}, J. Trujillo Bueno\altaffilmark{1,2,3} } 
\altaffiltext{1}{Instituto de Astrof\'{\i}sica de Canarias, 38205, La Laguna, Tenerife, Spain}
\altaffiltext{2}{Departamento de Astrof\'\i sica, Facultad de F\'\i sica, Universidad de La Laguna, Tenerife, Spain}
\altaffiltext{3}{Consejo Superior de Investigaciones Cient\'{\i}ficas, Spain} \email{ecarlin@iac.es}

\begin{abstract}
Velocity gradients in a stellar atmospheric plasma have an effect on the anisotropy of the radiation field that illuminates each point within the medium, and this may in principle influence the scattering line polarization that results from the induced atomic level polarization. Here we analyze the emergent linear polarization profiles of the Ca \textsc{ii} infrared triplet after solving the radiative transfer problem of scattering polarization in time-dependent hydrodynamical models of the solar chromosphere, taking into account the effect of the plasma macroscopic velocity on the atomic level polarization. We discuss the influence that the velocity and temperature shocks in the considered chromospheric models have on the temporal evolution of the scattering polarization signals of the Ca \textsc{ii} infrared lines, as well as on the temporally averaged profiles. Our results indicate that the increase of the linear polarization amplitudes caused by macroscopic velocity gradients may be significant in realistic situations. We also study the effect of the integration time, the micro-turbulent velocity and the photospheric dynamical conditions, and discuss the feasibility of observing with large-aperture telescopes the temporal variation of the scattering polarization profiles. Finally, we explore the possibility of using the differential Hanle effect 
in the IR triplet of Ca \textsc{ii} with the intention of avoiding the characterization of the zero-field 
polarization to infer magnetic fields in dynamic situations.
 
\end{abstract}


%%%%%%%%%%%%%%%%%%%%%%%%%%%%%%%%%%%%%%%%
%%%%%%%%%%%%%%%%%%%%%%%%%%%%%%%%%%%%%%%%
% INTRODUCTION
%%%%%%%%%%%%%%%%%%%%%%%%%%%%%%%%%%%%%%%%
%%%%%%%%%%%%%%%%%%%%%%%%%%%%%%%%%%%%%%%%
\section{Introduction} 

The chromosphere, the interface region between the photosphere and the corona,
is a very important part of the solar atmosphere. It is the place where most of the non-thermal energy that creates the corona and solar wind is released, with a heating rate requirement that is between one and two orders of magnitude larger than in the corona. To infer the thermal, dynamic and magnetic structure of the solar chromosphere is thus a very important goal in astrophysics. For instance, it is believed that the dissipation of magnetic energy in the $10^6$ K corona may be significantly modulated by the strength and structure of the magnetic field in the chromosphere \citep[e.g.,][]{parker07}. However, ``measuring" the chromospheric magnetic field is notoriously difficult \citep[e.g., reviews by][]{casini_landi07,harvey09,trujillo10}. While spectroscopic observations allow us to determine temperatures, flows and waves, they do not provide any quantitative information on the chromospheric magnetic field. To this end, we need to measure and interpret the polarization t
 hat some physical mechanisms introduce in chromospheric spectral lines. These mechanisms are the Zeeman effect, scattering processes and the Hanle effect.

The circular and linear polarization signals that the Zeeman effect can in principle produce in a spectral line
are caused by the wavelength shifts between the $\pi$ and $\sigma$ transitions of the line, as a result of the Zeeman
splitting induced by the presence of a magnetic field. The amplitude of the circular polarization scales with the ratio, $\cal R$, between the Zeeman splitting and the Doppler line width. The amplitude of the linear polarization scales with ${\cal R}^2$ \citep[see][]{landi_landolfi04}. 
%We note that ${\cal R}={{1.4{\times}10^{-7}{\lambda}B}{/}{\sqrt{1.663{\times}10^{-2}T/\alpha+{\xi}^2}}}$, where $\lambda$ is the spectral line wavelength in \AA, $T$ the kinetic temperature in K, $\xi$ the microturbulent velocity in ${\rm km}\,{\rm s}^{-1}$, $\alpha$ the atomic weight of the atom under consideration, and $B$ is in gauss 
Outside sunspots (where $B{\lesssim}100$ G at chromospheric heights) ${\cal R}{\ll}1$, which explains why it is so difficult to detect the polarization of the Zeeman effect in a chromospheric line. Typically, only the circular polarization is detected, especially in long-wavelength chromospheric lines such as those of the IR triplet of Ca {\sc ii} \citep[e.g., ][figure 3]{trujillo10}. But the linear polarization observed in quiet regions of the solar chromosphere has  practically nothing to do with the transverse Zeeman effect.

In weakly magnetized regions, the linear polarization of chromospheric lines is dominated by scattering processes.
The physical origin of this polarization is the difference among the electronic populations of sublevels pertaining to the levels of the spectral line under consideration. This  so-called atomic level polarization, which is caused by the anisotropic illumination of the atoms, produces selective emission and/or selective absorption of polarization components without the need of a magnetic field \citep[e.g.,][]{manso_trujillo03b,manso10}. The larger the anisotropy of the incident radiation field the larger the induced atomic level polarization and the larger the amplitude of the linear polarization of the emergent spectral line radiation. In an optically thick plasma like the solar atmosphere, the anisotropy of the radiation field depends mainly on the spatial distribution of the physical quantities that determine, at each point within the medium, the angular variation of the incident intensity. Great attention has been paid to the gradient of the source function \citep[e.g.,][]
 {trujillo01,landi_landolfi04} but, in a highly dynamic medium like the solar chromosphere, the gradients of the macroscopic velocity of the plasma may also play an important role \citep[][and references therein]{carlin12}. In fact, in \citet[][hereafter Paper {\sc i}]{carlin12} we showed that it can affect significantly the scattering polarization of the IR triplet of Ca {\sc ii}. Our arguments were based on radiative transfer calculations in a semi-empirical model of the solar atmosphere, after introducing ad-hoc velocity gradients and comparing the computed $Q/I$ profiles with those corresponding to the static case. Given the diagnostic potential of the Ca {\sc ii} IR triplet for exploring the magnetism of the solar chromosphere \citep[e.g.,][]{manso10,delacruz12}, and the fact that the region where such chromospheric lines originate may be affected by vigorous and repetitive shock waves \citep[e.g., ][]{carlsson_stein97}, it is necessary to investigate the radiative trans
 fer problem of scattering polarization in the Ca {\sc ii} IR triplet using dynamical, time-dependent atmospheric models of the solar chromosphere. In this paper, we show the results of such investigation.

%%%%%%%%%%%%%%%%%%%%%%%%%%%%%%%%%%%%%%%%
%%%%%%%%%%%%%%%%%%%%%%%%%%%%%%%%%%%%%%%%
% BASIC EQUATIONS
%%%%%%%%%%%%%%%%%%%%%%%%%%%%%%%%%%%%%%%%
%%%%%%%%%%%%%%%%%%%%%%%%%%%%%%%%%%%%%%%%
\section{Description of the problem and the resolution procedure.}\label{sec:cal} 
We have carried out radiative transfer calculations of the linear polarization produced by scattering in the  
Ca \textsc{ii} infrared (IR) triplet. The polarization is produced by the atomic level polarization that 
results from anisotropic radiation pumping in the hydrodynamical (HD)
models of solar chromospheric dynamics described in \cite{carlsson_stein97,carlsson_stein02}.

We used two time series of snapshots from the above-mentioned radiation HD simulations, each one lasting about 3600 s and showing the upward propagation of acoustic wave trains growing in amplitude with height until they eventually produce shocks. The first one corresponds to a relatively strong photospheric disturbance showing well-developed cool phases and pronounced hot zones at chromospheric heights (see \citeauthor*{carlsson_stein97}, \citeyear{carlsson_stein97}; we refer to this as the strongly dynamic case). The second simulation corresponds to a less intense photospheric disturbance (see \citeauthor*{carlsson_stein02}, \citeyear{carlsson_stein02}; we refer to this as the weakly dynamic case).  Thus, the thermodynamical evolution of the atmosphere (including the chromosphere and the transition region) 
is driven by the bottom boundary condition that is imposed on the velocity. This realistic boundary condition is extracted from the measured
Doppler shifts in the Fe {\sc i} line at $3966.8$ \AA. Our description focuses mainly on the strongly dynamic case, but in Sec. \ref{subsec:dynamic} we 
compare the results with those corresponding to the weakly dynamic case. 

To characterize the simulations we can use the following quantities. In terms of the velocity gradients, and using units related to a representative scale height  \footnote{A  scale height can be defined as the typical distance over which atmospheric magnitudes such as the density vary an order of magnitude. But the scale height is not a fixed quantity in a time-dependent model that contains important temporal variations in those magnitudes. For this reason we have defined an averaged scale height as the \textit{representative} value used for the characterization of the velocity gradients.} $\mathcal{ H}=275 \,\rm{km}$,
the temporal average of the maximum velocity gradient along the atmosphere is $40 \,\rm{km\cdot s^{-1}}$ per scale height (or $145 \,\rm{m\cdot s^{-1}km^{-1}}$) in the strongly dynamic case and $13 \,\rm{km\cdot s^{-1}}$ per scale height (or $47 \,\rm{m\cdot s^{-1}km^{-1}}$) in the weakly dynamic case. Likewise, 
the temporal average of the minimum of temperature in the atmosphere is $\rm{3976 \, K}$ in the
strongly dynamic case, and $\rm{4292 \,K}$ in the weakly dynamic case.

At each time step of the HD simulation under consideration 
we use the corresponding one-dimensional stratifications of the vertical velocity, temperature and density to compute the emergent $I(\lambda)$ and $Q(\lambda)$ profiles through the application of the multilevel radiative transfer code of \cite{manso_trujillo03a,manso10}, after the generalization to the non-static case described in \citet[]{carlin12}. Specifically, we have solved jointly the radiative transfer (RT) equations for the Stokes $I$ and $Q$ parameters and 
the statistical equilibrium equations (SEE) 
for the atomic populations of each energy level and the population imbalances 
among its magnetic energy sublevels (equivalently, the multipolar tensor components of the atomic density matrix, $\rho^K_0 (J_i)$, with $J_i$ the angular momentum of each level $i$). This is the NLTE radiative transfer problem of the second kind \citep[see sections 7.2 and 7.13 in][]{landi_landolfi04}). Once the self-consistent solution of such equations is found at each height in the atmospheric model under consideration, we compute the coefficients of the emission vector and of the propagation matrix (see section 2.2 of Paper I) and solve the RT equations
for a line of sight (LOS) with $\mu=0.1$, where $\mu$ is the cosine of the heliocentric angle. This LOS has been chosen in order to simulate a close to the limb observation, such as that shown in figure 13 of  \citet[][]{stenflo00b}. To account for the macroscopic motions, we have introduced the Doppler effect in the calculation of the absorption and emission profiles for each wavelength and 
ray direction (Paper {\sc i}). The influence of the Doppler
effect on the SEE appears directly because the radiative rates depend on the radiation field 
tensor components. Likewise, the RTE is affected because the Doppler effect modifies the elements of the propagation matrix and of the emission vector. 

Given that the computations reported here 
are carried out in plane-parallel atmospheric models, it is
necessary to introduce a micro-turbulent velocity, that accounts for the Doppler 
shifts (inducing an effective line broadening) produced by moving fluid elements 
below the resolution element. In order to estimate a suitable value (assumed constant with height), we have calculated the emergent intensities at disk center and
compared them with those of the solar Kitt Peak FTS Spectral Atlas \citep{kurucz_atlas84}. A good
agreement is obtained with $3.5$ km s$^{-1}$.

\section{Description and characterization of the results.}\label{sec:descript_results}
\begin{figure*}[htb]
\epsscale{1.0}
\plotone{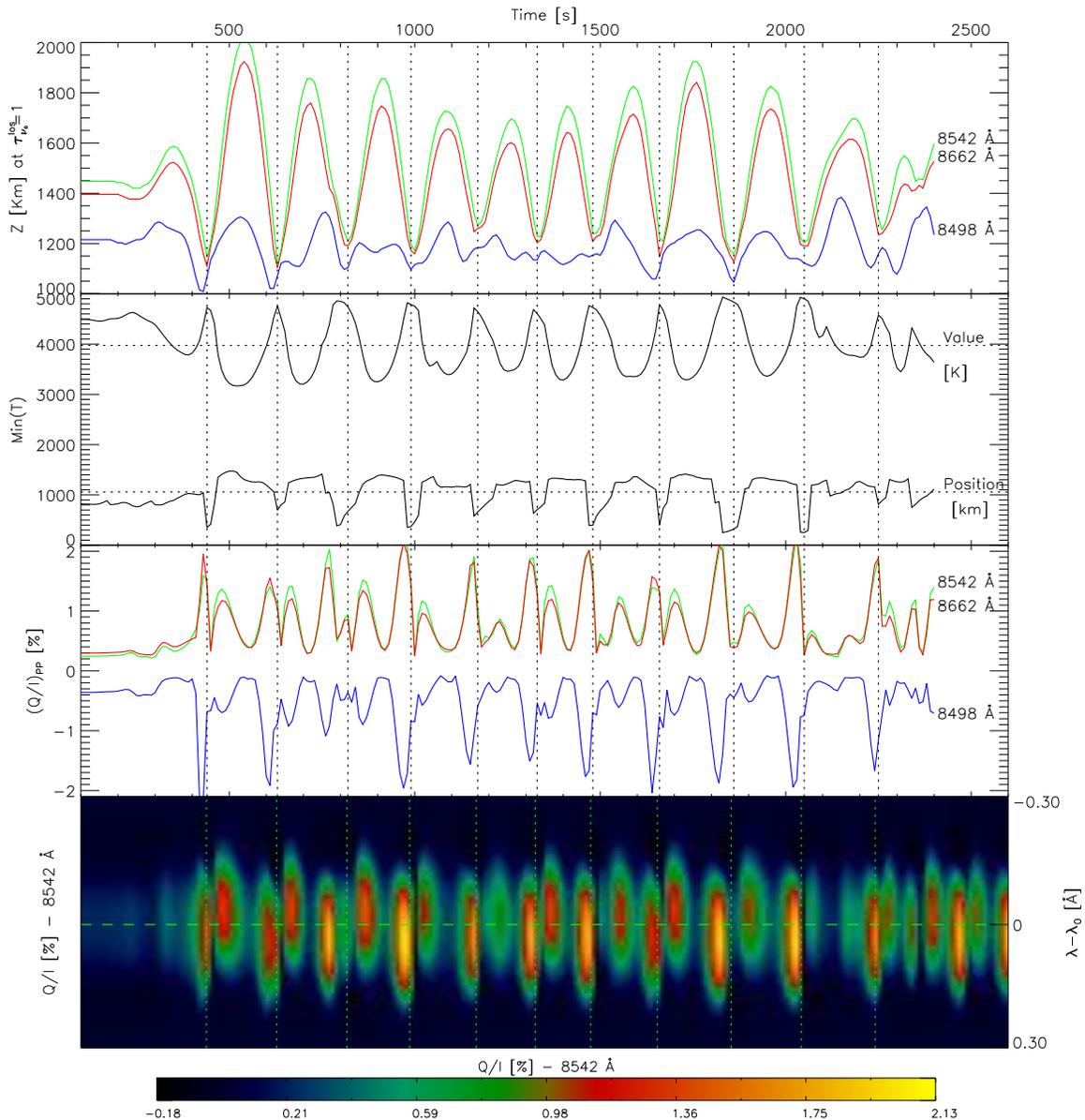}
%\plotone{tevol_paper_b.eps}
%\plotone{tevol_paper_c.eps}
\caption{Top row: time evolution of the atmospheric heights where $\tau^{los}_{\nu_0}=1$ for
the three Ca {\sc ii} IR transitions. The dotted vertical lines are located at the local minimum of
the $\tau^{los}_{8452}=1$ curve, and they can be considered to indicate the beginning and the end of each
``three-minute" period with their corresponding expansion and compression phases.
Second row: time evolution of the temperature value and atmospheric height of the temperature minimum.
Third row: time evolution of the polarization contrast
($\rm{\max(Q/I)-\min(Q/I)}$) for the three lines of the Ca {\sc ii} IR triplet. The polarization amplitude of the ${\lambda}8498$ 
line has been multiplied by -5 to show the results for the three lines on the same scale. Note that, by definition, the line contrast is
always positive, so with this quantity we cannot know if the polarization signal is positive or negative. Thus, with the artificial sign inversion done for the 8498 \AA\
line contrast in this figure, we try to illustrate that this line is the only one whose larger polarization amplitudes are negative.Bottom row: time evolution of the calculated $Q(\lambda)/I(\lambda)$ fractional linear polarization profile
of the ${\lambda}8542$ line.\label{fig:tevol_all}}
\end{figure*}

A standard Fourier analysis of the atmosphere model shows that it acts as a 
pass-band filter for the multifrequency sound waves generated in the lower boundary. The result 
is that the predominating periods at chromospheric heights and higher are around three minutes \citep{carlsson_stein97}. For practical reasons
we divided the temporal evolution in 3-min intervals so that the beginning of
each interval coincides with the moment in which the shock front in temperature and velocity 
is the sharpest in each interval (vertical lines in figures with temporal axis, like Fig. \ref{fig:tevol_all}). 
Given the power of the 3-min waves, this division turns out to be ``natural'' and can be used to
mark the most interesting events we see in the emergent polarization.

Inside each three-minutes cycle we distinguish between \textit{compression} and \textit{expansion} phases.
They can be easily identified following the height at which $\tau^{los}_{\nu_0}=1$, i.e., 
where the optical depth at line center ($\nu_0$) along the LOS equals unity (upper panel of Fig. \ref{fig:tevol_all}). This quantity
is a good marker of the shock fronts when they cross heights between $1$ and $2$ Mm. It is because the
steep changes in opacity inside the shocks forces the $\tau=1$ region to remain comprised within them.
The line transitions at $8542$ \AA\ and $8662$ \AA\ (green and red lines in the upper panel
of Fig. \ref{fig:tevol_all}) follow a clearer periodic pattern because they form higher, 
where less frequency components of the velocity waves can arrive.
Compression phases begin when plasma falls 
down from upper layers ($\tau^{los}_{8542}=1$ and $\tau^{los}_{8662}=1$ decrease in top panel 
of Fig. \ref{fig:tevol_all}), while simultaneously a new upward propagating wave emerges amplified
into the chromosphere. At the end of this stage a shock wave is completely developed and 
the $\tau^{los}_{\nu_0}=1$ position is close to $\sim$1200 km for the three IR lines. The shock waves
so created start always in such region between 1 and 1.5 Mm \footnote{It is in this range of heights  where the Ca {\sc ii} IR triplet forms in typical semi-empirical models}. After that, during what we term expansion phase 
(heights for $\tau^{los}_{8542}=1$ and $\tau^{los}_{8662}=1$ arising in top panel of Fig. \ref{fig:tevol_all}), the shock 
fronts travel upward increasing the plasma velocities as they encounter lower densities. 

\begin{figure*}[!t]
\epsscale{1.2}
\plotone{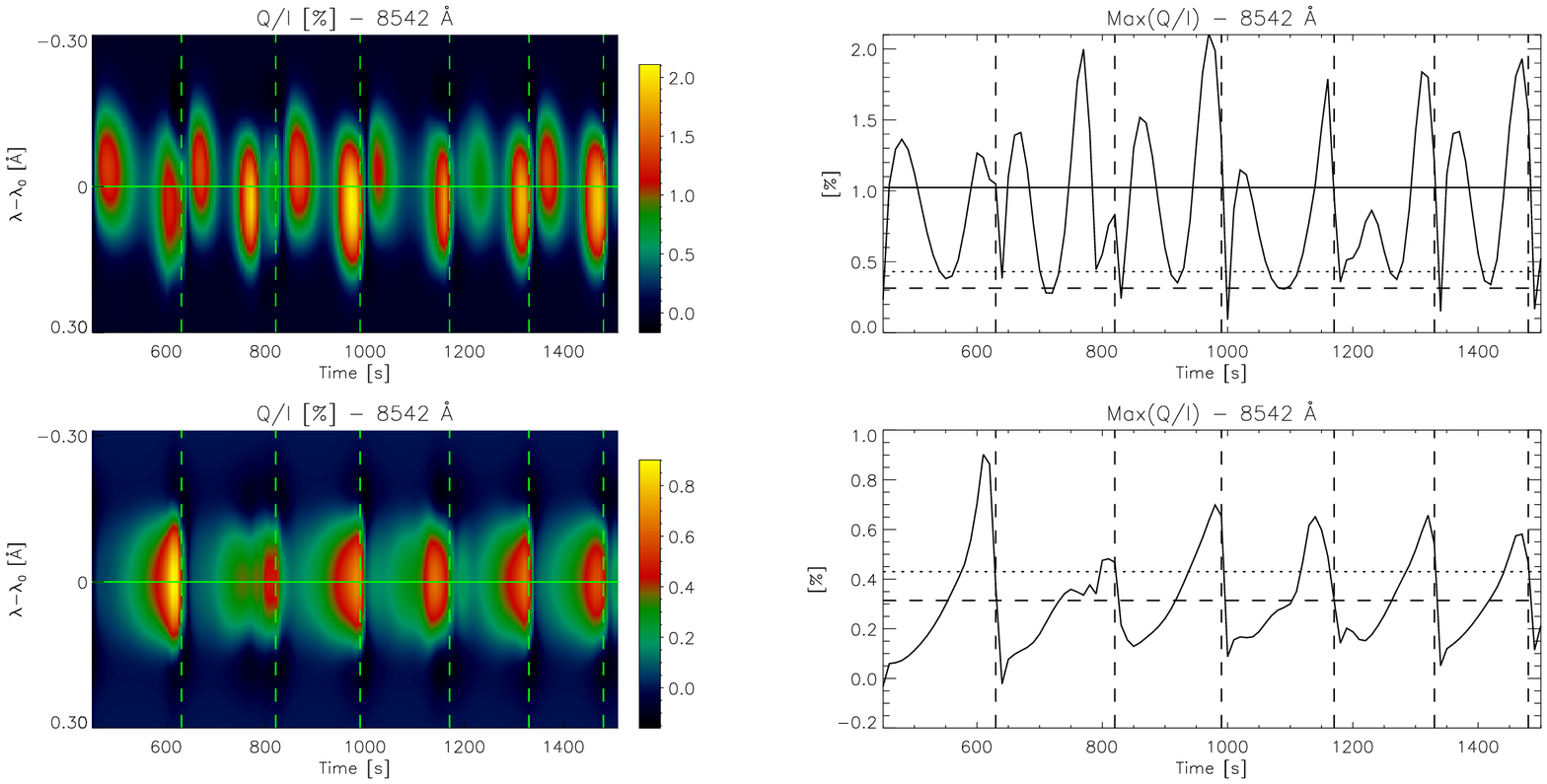}
\caption{Temporal evolution of $Q(\lambda)/I(\lambda)$ (left column panels) and of the 
${\rm{Max}}(Q/I)$ peak amplitude (right column panels) during 1040 s (17 minutes) for the 8542 \AA\ transition. 
The results of the upper panels take into account the effect of the velocity gradients, while those of the lower panels were obtained assuming zero macroscopic velocities while calculating the density-matrix elements. 
The vertical lines indicate the beginning and the end of each three-minutes period. 
The horizontal lines in the right column panels show the temporally-averaged amplitudes ($\langle Q \rangle/\langle I \rangle$) obtained from the $Q/I$ profiles computed in the strongly dynamic model,  
with the solid line indicating the polarization amplitude obtained when the effect of the velocity gradients is taken into account and the dashed lines showing the polarization amplitude obtained assuming zero macroscopic velocities. The dotted horizontal lines in the right panels indicate the polarization amplitudes obtained in the (static)   
FAL-C semi-empirical atmospheric model. \label{fig:qconysinv}}
\end{figure*}

Figure \ref{fig:tevol_all} also shows the time evolution of other quantities
during the first $2000$ s after the initial transient. In the second row, the
location and value of the temperature minimum are displayed, showing a clear correspondence with
expansion and contraction phases. In the third row, we show the ensuing
variation of (Q/I)$_\mathrm{pp}$, defined as the peak-to-peak difference of the
Q/I profile for each spectral line. It is a measure of the
linear polarization signal contrast that was used in Paper \textsc{i} to characterize the polarization amplitude and
discriminate their variations with respect to static cases. In each cycle we see
an amplification of (Q/I)$_\mathrm{pp}$ occurring at expansion phases and an
usually larger amplification during contraction phases. Finally, the time evolution
of the emergent Q$(\lambda)$/I$(\lambda)$ profile for the $8542$ \AA\ line is illustrated in the lower panel (the
vertical axis shows 0.6 \AA\ around the rest wavelength of the line).
Here, we observe two distinct areas showing amplifications inside each
three-minute cycle. The first amplification is blue-shifted, because it happens
in an atmospheric expansion phase (plasma moving towards the observer). It 
is weaker than the second amplification, which is red-shifted and occurs during the compression phase
(plasma moving down in the atmosphere).
This indicates that the compression phase is more efficient producing a
polarization amplification than the expansion one. The reason is that during compression we have stronger velocity 
and temperature gradients along the main
regions of formation. Following the results of Paper \textsc{i}, the
larger the gradient, the larger the enhancement of the linear polarization signal.
The behaviour is similar in the other transitions.

There is a clear correspondence between the maximum value of the temperature minimum
(hot-chromosphere time-steps) and the largest peaks of the $\rm{(Q/I)_{pp}}$ signal,
taking place just before the maximum contraction (dotted vertical lines). As the
atmosphere is compressed, the temperature increases at chromospheric heights and
the resulting gradient of the source function produces an increase of the radiation field 
anisotropy in the upper layers. This directly leads to an enhanced emergent linear polarization
signal. On the contrary, in cold-chromosphere models the expansion reaches its maximum and $\rm{(Q/I)_{pp}}$ is near its minimum value.

Even in such complex situations, we still witness the already known effects of amplification (with respect to the static case), frequency
shift and asymmetry in the linear polarization profiles due to dynamics. All of them have been
already explained in Paper
{\sc i}, using the semi-empirical FAL-C model of Fontenla et al. (1993) with ad-hoc velocity stratifications. The enhancement is produced as a consequence of the velocity gradients and subsequent anisotropy
enhancements. However, some differences exist from the experiments in semi-empirical models
and the calculations presented in this paper. First, the velocity stratification in
the HD models is, in general, non-monotonic and with a non-constant variation with height. Second,
the maximum velocity gradients are located at shocks, with amplitudes that reach tens or
even hundreds of meters per second per kilometer (as a comparison, in Paper {\sc
i} we dealt with velocity gradients between 0 and 20 m s$^{-1}$ km$^{-1}$). 
Third, as commented before, we have shocks in temperature that produce larger
source function gradients and additional enhancement of the
radiation anisotropy and of the linear polarization. Finally, these variations are
usually concentrated in the formation regions of the triplet lines. 
 All these mechanisms act together and  
enhance the linear polarization of the emergent radiation with amplification
factors up to $\sim10$ (in the $8498$ \AA line) and $\sim7$ (in the $8542$ \AA\ and $8662$ \AA\
lines), for the instantaneous values of the $\rm{Q/I}$ amplitudes with respect to
the static FAL-C case. However, if we consider temporal averages of the emergent Stokes profiles during long periods, we get amplification factors of about a factor
of 2 (time-averaged Q/I amplitudes reach $\sim1 \,\%$ for 8542 \AA\ and 8662 \AA\ lines). 

\begin{figure*}[!t]
\epsscale{0.3}
\plotone{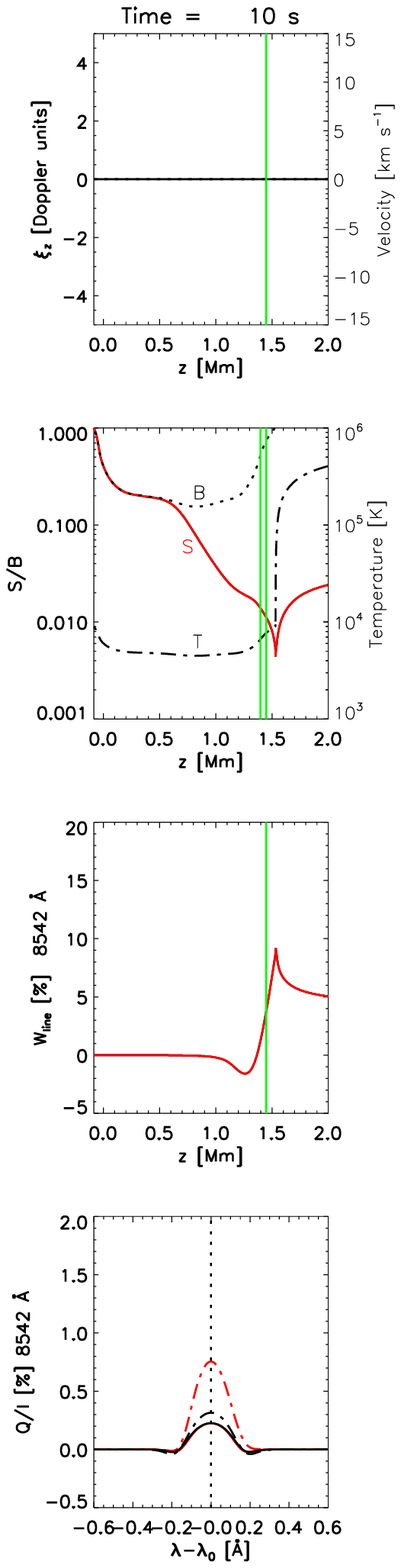}
\plotone{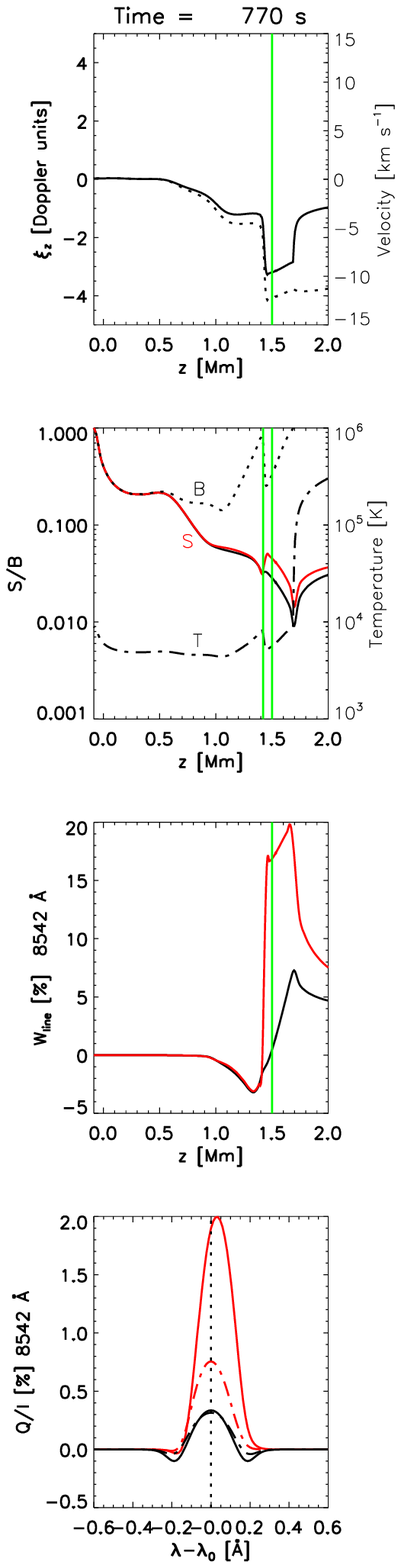}
\plotone{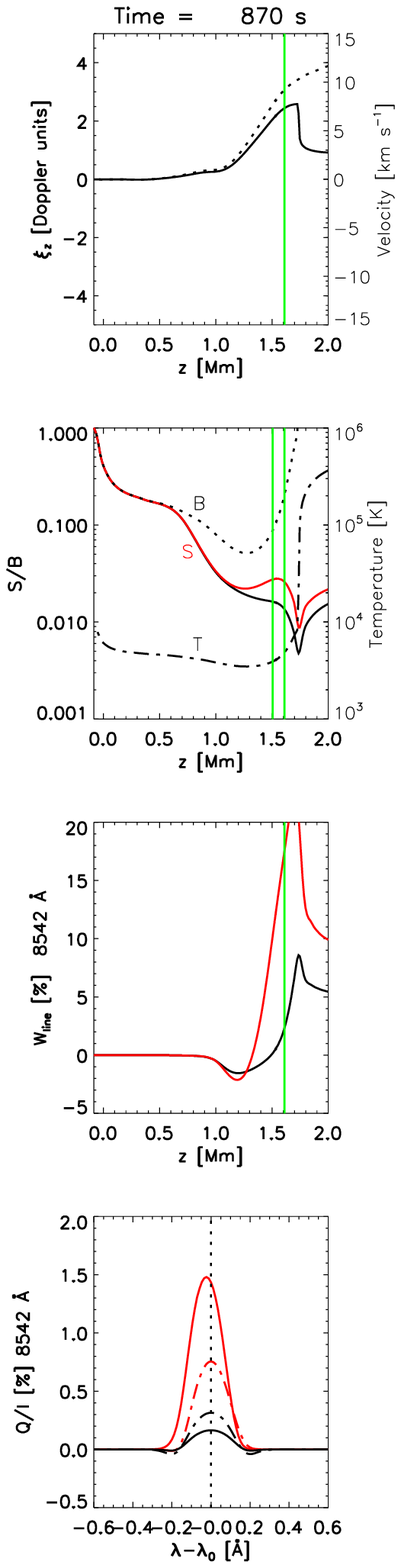}
\caption{This figure shows the values of the quantities indicated in the vertical axes 
for the time steps 10 s. (left column, a `quiet'
situation), 770 s. (middle column, a compression stage) and 870 s. (right
column, an expansion phase) of the strongly dynamic hydrodynamical simulation. 
First row (from top): macroscopic vertical velocity (dotted
line) and adimensional vertical velocity $\xi_z$ (solid line) versus height. The green
line of each panel indicates the atmospheric height where $\tau^{los}_{\nu_0}=1$. Second row (from
top): temperature (dashed line), Planck function (dotted line), source function
for the zero velocity approximation (solid black line) and source function
allowing the influence of the model's velocity gradients (red solid line). The green lines mark the
instantaneous positions of $\tau^{los}_{\nu_0}=1$ and $\tau^{los}_{\nu_0}=2$.
Third row (from top): line anisotropy factor (see equation 4 in Carlin et al. 2012) calculated for each of the above-mentioned cases; neglecting the effect of the velocity gradients (black solid line) 
or allowing it (red solid lines).
Fourth row (from top): emergent $\rm{Q/I}$ profiles versus wavelength with respect
to the line center, with the same color code than in previous panels. The dashed lines
here are the time averages over the entire simulation, for each case (with or without velocities).\label{fig:termo}}
\end{figure*}

Summarizing, the temporal evolution of the polarization is driven by the temperature and velocity
stratifications, that in turn are a result of the dynamical conditions set in the
photosphere.

%To characterize the simulations we can use the following quantities. In terms of the velocity gradients, and using units related to a representative {\bf scale height}  \footnote{A  scale height can be defined as the typical distance in which atmospheric magnitudes such as the density vary an order of magnitude. But the scale height is not a fixed quantity in a time-dependent model that contains important temporal variations in those magnitudes. {\bf For that reason we have defined an averaged scale height as the \textit{representative} value} used for the characterization of the velocity gradients.} {\bf $\mathcal{ H}=275 \,\rm{km}$},
%the temporal average of the maximum velocity gradient along the atmosphere is 
%%$145 \, \pm \,172\,\rm{m\cdot s^{-1}km^{-1}}$ for the strongly dynamic case
%%and $47 \, \pm \,77 \,\rm{m\cdot s^{-1}km^{-1}}$ for the weakly dynamic case. 
%$40 \, \pm \,47.3\,\rm{km\cdot s^{-1}}$ per scale height for the strongly dynamic case
%and $13 \, \pm \,21.1 \,\rm{km\cdot s^{-1}}$ per scale height for the weakly dynamic case. The numbers following the $\pm$ signs are the standard deviation of the maximum velocity gradient, what as usual gives a measure of the distribution of the instantaneous values around the mean. Likewise, 
%the temporal average of the minimum of temperature in the atmosphere is $\rm{3976 \,\pm\,507 \, K}$ in the
%strongly dynamic case, and $\rm{4292 \,\pm\,379 \,K}$ in the weakly dynamic case.

\section{Analysis and discussion of results.}\label{sec:results}

\subsection{The effect of the velocity.}\label{subsec:velocity}
A way of visualizing the effect of vertical velocity gradients on the emergent scattering polarization
 is to compare the evolution of the polarization profiles corresponding to both the static and non-static
case. In the absence of velocities (lower row of Fig. \ref{fig:qconysinv}), the
maximum of the $Q/I$ profiles is always located at $\lambda=\lambda_0$ (i.e., line center), and its
temporal evolution presents a sawtooth shape.
When the effect of velocities is included in the calculations (upper row
of Fig. \ref{fig:qconysinv}), the maximum of the $Q/I$ signal is no longer located at the
central wavelength and its temporal evolution assumes a different shape with two
peaks every 3-min period (upper right panel). These wavelength and amplitude modulations are produced by the Doppler effect of the velocity gradients.

It is interesting to compare the mean $\rm{Q/I}$ amplitudes obtained in the hydrodynamical models
with the one calculated in the FAL-C model. They differ notably (see horizontal lines in right panels of Fig. \ref{fig:qconysinv}). 
In the 8542 \AA\ transition we have mean values around 1\%, 0.31\% and 0.42\% for the HD models with velocities, the HD
models at rest and static FALC models, respectively. The results of these figures have been obtained using an
integration time of 1040 s ($\backsim 17$ minutes), the duration of the temporal interval
shown in Fig. \ref{fig:qconysinv}.
Neglecting the effect of the velocity gradients in the HD models, we see that the 
resulting temporally-averaged scattering polarization signals (which include the impact of the temperature and density shocks) are similar to
the $Q/I$ profiles computed in the static FAL-C semi-empirical model.

\subsection{The combined effect of velocity and temperature on the linear polarization.}\label{subsec:combined}
In Fig. \ref{fig:termo} we display some relevant magnitudes for three different
situations in the simulation. The first column corresponds to a \textit{quiet} time-step, with no shocks, zero
velocity and without any kind of amplifications (it is the initial transient phase).
The central column shows a phase of \textit{compression}, in which shocks are important. Finally,
the last column displays an \textit{expansion} phase, in which the atmosphere is
expanded and the shocks are already travelling over the transition region. 
Furthermore, we distinguish between the solutions when motions are
taken into account (red lines) and the solutions obtained allowing shocks in all magnitudes
but artificially setting the velocity to zero (black lines).

The normalized velocity $\xi_z=(\nu_0/c)v_z/\Delta\nu_D$, with $\Delta\nu_D$
the Doppler width of the absorption profiles (that depends on the temperature), $c$ the speed
of light and $v_z$ the vertical velocity, is the quantity that controls
the importance of the atmospheric motions in relation with the radiation anisotropy and the scattering polarization (see Paper {\sc i}). Note that this quantity considers the combined effect of velocity and temperature. In the HD atmosphere models, $\xi_z$ (solid lines in upper panels of Fig.
\ref{fig:termo}) is only significant in the formation region of the IR triplet lines
($\tau^{los}_{\nu_0} \sim 1$ region with high velocity gradients, and not very high temperatures). Although shock waves increase the
chromospheric temperature, the effect of the velocity gradients is predominant. The
contrary occurs over the transition region, where the thermal line width is much larger than the 
Doppler shifts.

%and consequently the effects of motions on the line transfer are negligible.

The expansion and contraction can be identified
also in quantities such as the intensity source function and the
Planck function (second row in Fig. \ref{fig:termo}). During contraction phases (middle
column panels), high temperatures produce a more efficient population pumping
towards upper levels, incrementing the emissivity and, consequently,
the source function. Additionally, during contraction the temperature shock occurs in
optically thick and denser layers (deeper layers below $\tau^{los}_{\nu_0}=1$), forcing the source function gradient to increase with respect to the static case at those
heights. Note how in this last case the source function rises \textit{as a whole} because of the warming (compare the source function in the middle panel, the black solid line that has been obtained neglecting velocities, with the non-dynamic source function in the left column). If the macroscopic velocity is now considered, we additionally get a jump in the source function (red lines in middle column of Fig. \ref{fig:termo}) caused by the velocity shock that is developed in this contraction phase. This behaviour is accompanied by a significant Doppler-induced anisotropy enhancement that amplifies the linear polarization, as shown in the corresponding lower panels of the same figure.  

In the expansion phases, the shock waves move upward and the
chromosphere becomes cooler. This induces a lower source function and smaller
polarization amplitudes (as compared with the contraction phase). Otherwise, as
the density of scatterers is now lower around the shock (because it moved upward
to regions with $\tau^{los}_{\nu_0}<1$), the temperature gradients have
smaller effects on the polarization profiles than during the contraction phases. In this expansion 
time step, the black solid line representing the \textit{static} source function is similar to the non-dynamic source function of the left column. However, once the motions are introduced, and despite of the fact that the shocks have already reached upper chromospheric layers, the remanent velocity field has still a sizable gradient that enhances the source function (Doppler brightening effect).

\begin{figure}[!t]%[htb]
\epsscale{0.8}
\plotone{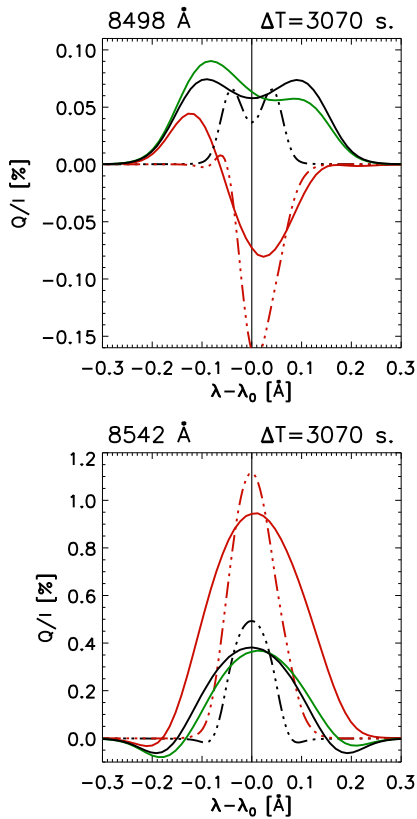}
\caption{Fractional linear polarization profiles of the 8498 \AA\ and
8542 \AA\ lines after temporally averaging the Stokes $I$ and $Q$ profiles 
during 3070 seconds (51 minutes). These $\langle Q \rangle/\langle I \rangle$ 
profiles may be considered to emulate what can be actually observed with today's solar telescopes.
Black solid profiles: static case with
$v_{\rm{micro}}=3.5\,\rm{km\,s^{-1}}$. Red solid profiles: strongly dynamic case 
taking into account the effect of the velocity gradients and assuming 
$v_{\rm{micro}}=3.5\,\rm{km\,s^{-1}}$. Black dashed profiles: strongly dynamic case 
neglecting the effect of the velocity gradients and assuming
$v_{\rm{micro}}=0$.Red dashed profiles:  strongly dynamic case 
taking into account the effect of the velocity gradients and assuming $v_{\rm{micro}}=0$.
The green solid lines show the temporally averaged profiles 
obtained after applying the velocity free approximation (VFA) with $v_{\rm{micro}}=3.5\,\rm{km\,s^{-1}}$
(i.e., neglecting the Doppler shifts of the macroscopic velocities when computing the density matrix elements, but taking them into account when calculating the emergent Stokes profiles). 
\label{fig:qprom}}
\end{figure}

\subsection{Averaged values of the polarization profiles.}\label{subsec:velo2}
In order to compute the average linear polarization signal that one would
observe without any temporal resolution, we average 
$Q$ and $I$ (obtaining $\langle Q \rangle / \langle I \rangle$) over $3070\,\rm{s}$ ($\approx 51$ minutes) for four
different cases (Fig. \ref{fig:qprom}). We consider the cases with zero 
micro-turbulent velocity (dotted lines)  and a constant
micro-turbulent velocity of 3.5 km s$^{-1}$ (solid lines). For each
case, we distinguish between the results \textit{switching off} the velocity
(black lines) and the results allowing for macroscopic velocity fields (red
lines). 

When macroscopic motions are considered, the polarization profiles become
asymmetric. Furthermore, they become more negative in the case of the
8498 \AA\ transition and more positive in the other two
transitions. The asymmetry of the red profiles is a consequence of the fact that, during the
averaging period, the dynamical situations in which the velocity gradient is
negative (velocity field mostly decreasing with height) dominate over the
situations with velocity gradients that are mostly positive. This predominance
is not because the situations with negative velocity gradients are more frequent but because such
situations are more efficient on amplifying the linear polarization. This
happens during the compression phase because i) the velocity gradients are larger, ii)
there is also a shock in temperature affecting the formation region and iii)
the shock fronts are located just below the $\tau^{los}_{\nu_0}=1$ height.
The results are qualitatively the same independently of the micro-turbulent
velocity value, but, when it is not considered, the amplification of
$\rm{(Q/I)_{pp}}$ is larger and the profiles are narrower. 

If we decrease the averaging interval to $9$ minutes, we obtain profiles that are essentially
similar to the ones obtained by averaging during $51$ minutes (showed in Fig. \ref{fig:qprom}). If we
integrate less than that, significant variations appear in the shape and
amplitude of the emergent profiles. This indicates that, concerning the linear polarization, there
is still reliable dynamic information contained in a time interval corresponding
to a few 3-min cycles.

%This depends on the temporal coherence and periodicity of the bottom boundary conditions of this particular situation.

\begin{figure*}[!h]
\epsscale{1.0}
\plotone{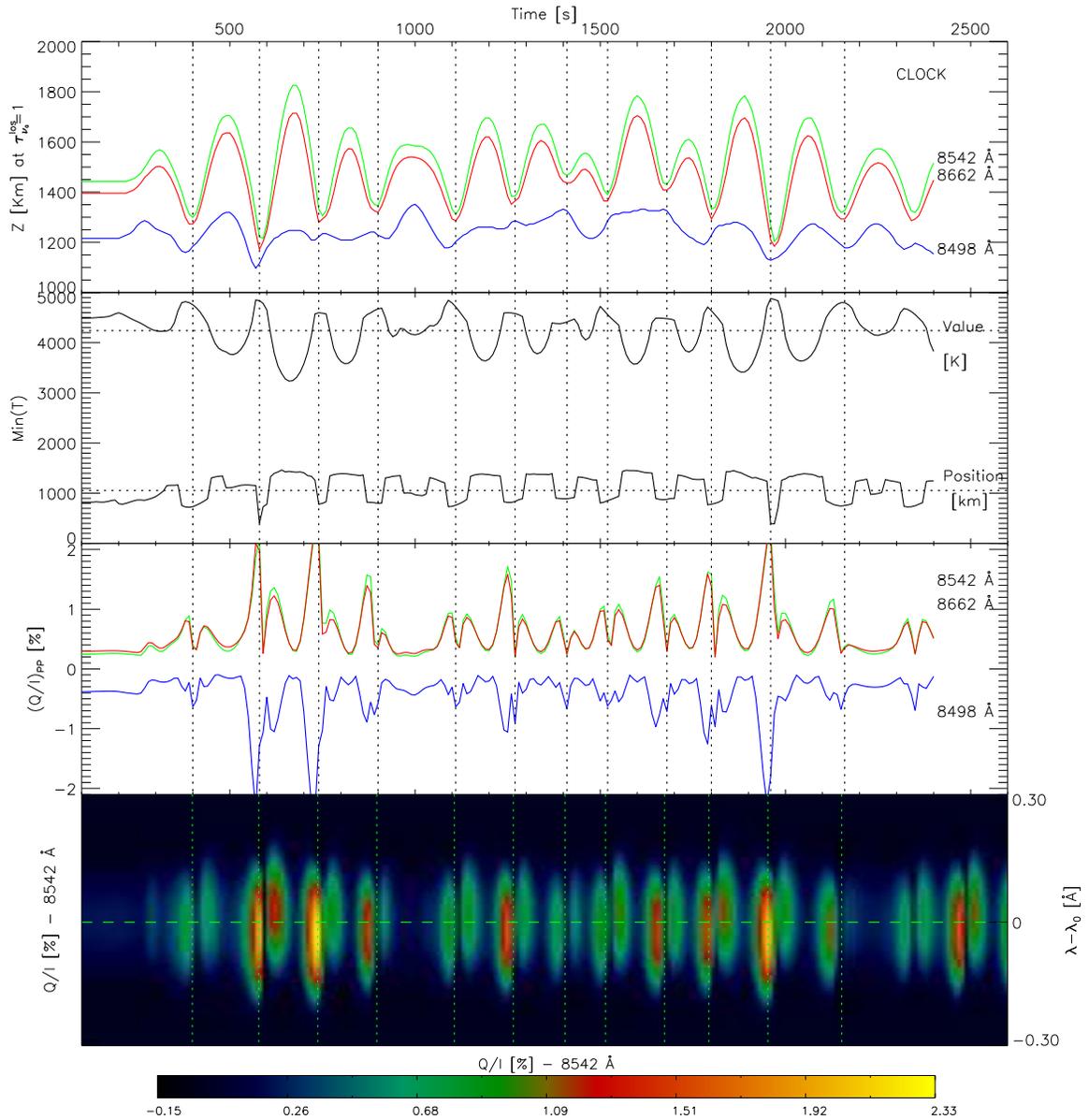}
\caption{ Same as Fig. \ref{fig:tevol_all}, but for the weakly dynamic
case. Remember that the blue line amplitude has been multiplied by -5 for scale
reasons. \label{fig:tevol_all_weak}}
\end{figure*}

\begin{figure*}
\epsscale{1.2}
\plotone{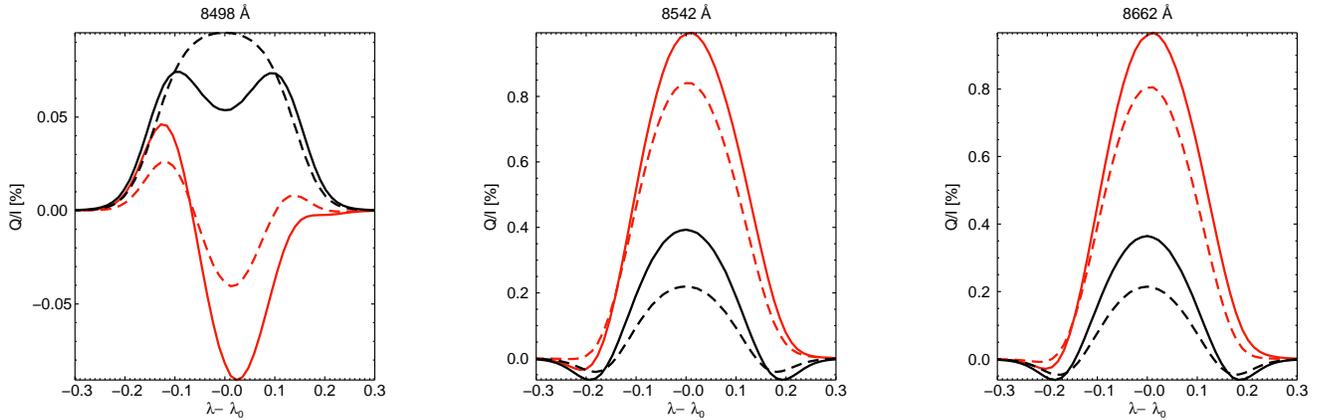}
\caption{Comparison of the results in the strongly and weakly dynamic cases (see
Sec. \ref{subsec:dynamic}). All profiles are the resulting $\langle Q \rangle/ \langle I \rangle$
profiles obtained averaging $Q$ and $I$ during 15 minutes of the considered simulation. Solid lines: results calculated in the strongly dynamic
case. Dashed lines: results calculated in the weakly dynamic case. Black lines: results allowing variations in all magnitudes but neglecting the velocity. Red lines: results allowing the effect of the velocity gradients. \label{fig:weak}}
\end{figure*}

\begin{figure*}[!t]
\epsscale{0.8}
\plotone{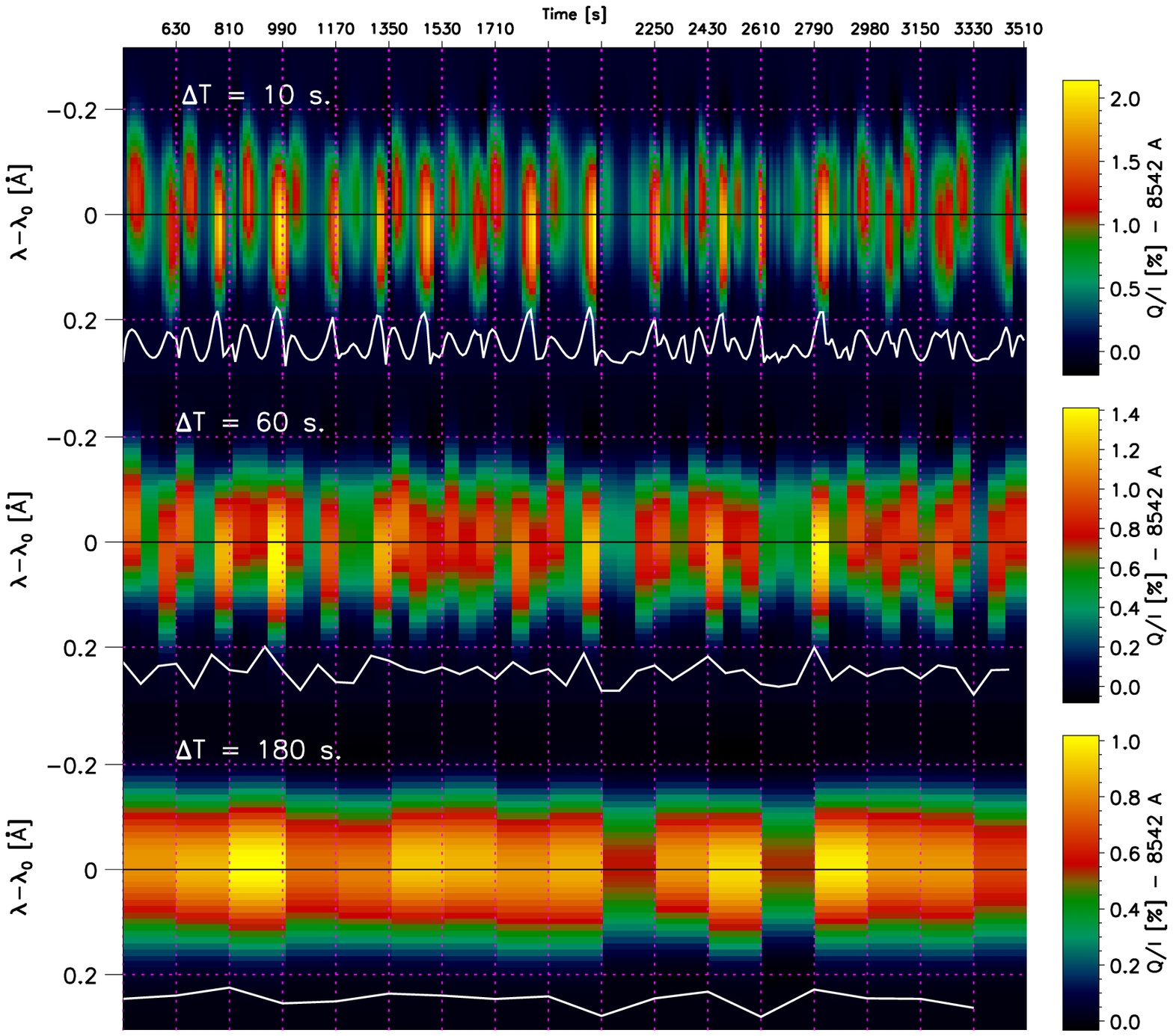}
\caption{Temporal evolution of the emergent $8542\lambda$ $Q/I$ profiles
for different integration times. Wavelength is in the vertical axis. From top to
bottom we have a 10, 60 and 180 s. temporal resolution, respectively. The white solid lines show
the temporal evolution of $\rm{(Q/I)_{pp}}$ (i.e., the amplitude contrast at each time-step). Vertical dotted lines mark each
three-minute period. \label{fig:int_effect}}
\end{figure*}

\subsection{The velocity free approximation.}\label{subsec:velo3}
An approximation that is sometimes applied to solve radiative transfer problems 
in dynamical atmospheres
 (either taking into account the presence of atomic polarization or not)
is the velocity free approximation (VFA). It is based on solving the SEE and
RTE simultaneously but neglecting the effect of plasma motions. 
However, once they are consistently
solved, such plasma motions are included in the synthesis of the
emergent Stokes profiles (along $\mu=0.1$ in our case). Consequently, the density matrix elements are calculated as if plasma motions did not affect them, reducing the 
complexity and computational effort of the problem since a reduced frequency grid is
used to compute the mean intensity and the anisotropy.
The results of applying it to each time step of our HD evolution is the temporal
average illustrated as the green line in Fig. \ref{fig:qprom}. This approximation is
clearly not appropriate in our case, given that the profiles just become asymmetric (with
respect to the static profiles) but without the amplification. The reason
for this lack of amplification is that the anisotropy controlling the linear
polarization is not correctly enhanced (see Paper {\sc i}). On the other hand, the asymmetry is
purely due to the asymmetric absorption with respect to the line center that
motions produce along the ray under consideration. Hence, in order to obtain reliable results 
it is mandatory to include the effect of
Doppler shifts in the whole set of equations, and we conclude that the VFA should not
be applied.

\begin{figure}[!h]
\epsscale{0.95}
\plotone{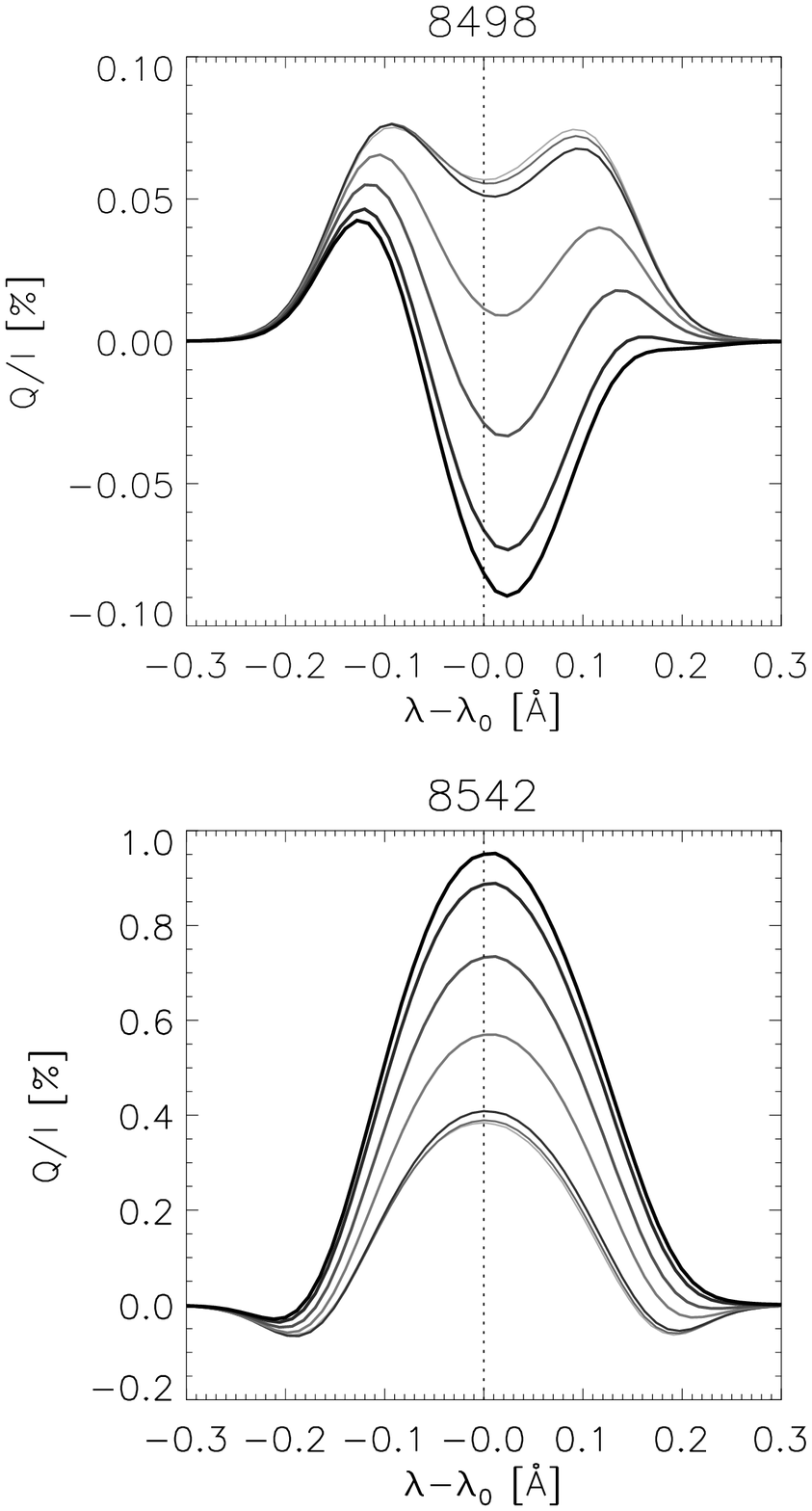}
\caption{Resulting $\rm{Q/I}$ profiles after averaging Q and I during $15$ minutes in the strongly dynamic simulation, using different values of $F$. $F$ is the scaling factor by which we have multiplied the modulus of the macroscopic velocity at each atmospheric height. The curves correspond to $F=1,\,0.9,\,0.7,\,0.5,\,0.2,\,0.1,\, 0$, going from black colour (fully dynamic case) to lighter grey colour (static case). The results for the 8662 \AA\ line are very similar to the ones obtained for the 8542 \AA\ line. \label{fig:consistency_test}}
\end{figure}

\subsection{The effect of photospheric dynamics.}\label{subsec:dynamic}
Given that the small velocity fields appearing in the photosphere are amplified because of the exponential decrease in the density while
the perturbations travel outwards, the properties of the bottom boundary condition are determinant in
the behaviour of the emergent Stokes parameter of chromospheric lines. We
compare the strongly dynamic case that forms the core of our paper with the 
weakly dynamic case that has been already introduced in Sec.
\ref{sec:descript_results}. Although the mean maximum velocity gradient is three times smaller in the
weakly dynamic case and the averaged polarization amplitudes are also
smaller than in the strongly dynamic one, we still find comparable or even slightly larger instantaneous $\rm{(Q/I)_{pp}}$ amplitudes (see Fig. \ref{fig:tevol_all_weak}).
The resulting averaged polarization profiles are
qualitatively the same but they differ in amplitude (Fig. \ref{fig:weak}).
This is a reasonable result because in the weakly dynamic scenario the instantaneous velocity
gradients are smaller in general.
Differences are especially critical for the 8498 \AA\ line, whose linear
polarization profiles can be positive, but also adopt significant negative
values at redder wavelengths (when velocity gradients are mainly positive with
height) or at bluer wavelengths (when velocity gradients are mainly negative
with height). This behaviour produces cancellation effects with integration
times larger than a three-minute period.
Furthermore, the central depression produced in the 8498 \AA\ average profile when the velocity is neglected in the strongly dynamic simulation (solid black line in left panel of Fig. \ref{fig:weak}) do not appear in the average profile corresponding to the weakly dynamic case (dashed black line in the same panel) because of the differences in the instantaneous temperature stratifications. The sensitivity of this spectral line to the instantaneous photospheric perturbations and to the developed chromospheric shocks is larger than in the other two lines.

\begin{figure*}[!t]
\epsscale{1.0}
\plotone{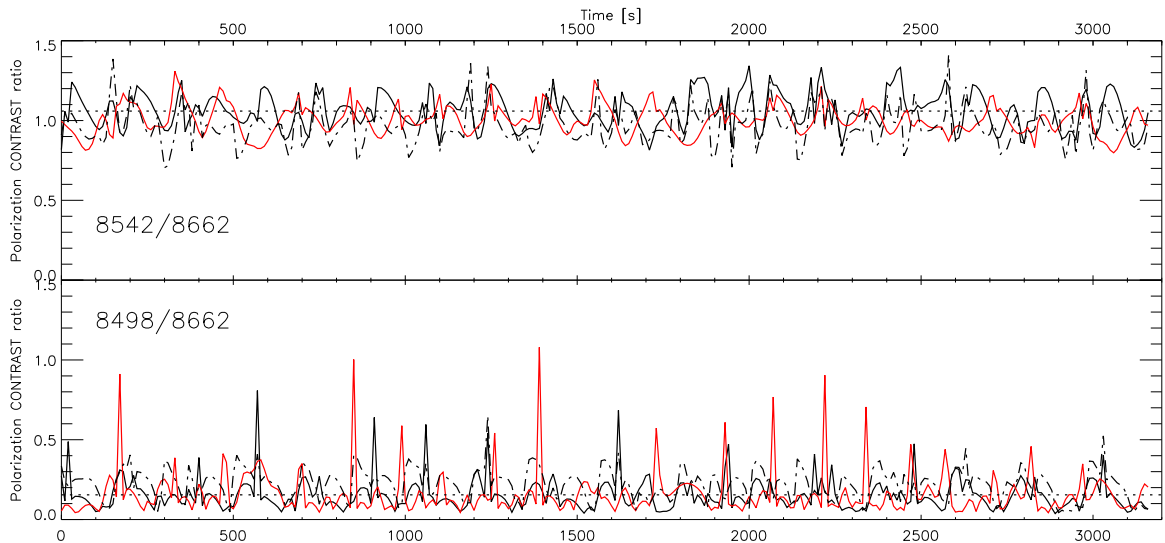}
\caption{ Black solid lines correspond to the strongly dynamic case. Red lines
correspond to weakly dynamic case. Black dashed lines correspond to the strongly
dynamic case but doing the calculations without micro-turbulent velocity. The
line ratio  $\varrho_3 = (\rm{Q/I})^{8498}_{pp}/(\rm{Q/I})^{8542}_{pp}$ is not
shown because it is very similar to $\varrho_1$ and because can be obtained from
the other two ratios. \label{fig:lineratios}}
\end{figure*}

\subsection{The effect of the integration time.}\label{subsec:tint}
 In order to detect in the Sun the time evolution of the linear polarization signals, the observations must have enough time resolution, signal to noise
ratio and spatial resolution. A sufficient spatial coherence is important to avoid
cancellations of the contribution from different regions in the chromosphere evolving with different
phases. If we consider the expected capabilities of the next generation of solar
telescopes (like the European Solar Telescope, EST, or the Advanced Technology
solar Telescope, ATST), we can aim at observing the
emergent Stokes profiles of Fig. \ref{fig:int_effect} with a 10 s cadence (upper panel)\footnote{Using EST (telescope diameter of 4 m, instrumental efficiency around 10\%) and
considering a spectral resolution of 30 m\AA, a spatial resolution of $0.1$
arcsecs and an integration time of 1 s (ten times better than needed), it would
be possible to observe the linear polarization of the 8542 \AA\ line (line to
continuum ratio of $\sim0.2$) at the level of $Q/I \sim10^{-3}$ with a confidence
of $3 \sigma$ over the noise.}. However, with the present telescopes and 
instrumentation, we are forced 
to integrate in time and/or space to detect 
the scattering polarization signals. If we degrade the temporal resolution of our results to an
integration time of 1 min (middle panel of Fig. \ref{fig:int_effect}) and 3
min (lower panel of  Fig. \ref{fig:int_effect}) we clearly see that the
time evolution becomes more difficult to detect. In
the last case, the profiles are already so smoothed that the
original features are completely lost, both in the spectral and temporal
domains. The amplitude of the integrated signals are lower than in the
original 10 s sequence by a factor of $2$ (see the color scales). However,
integration during time intervals ${\sim}1$ min could reveal the
amplification/modulation effect if we capture spectro-polarimetric signals
similar to the ones showed in the middle row panel.

\subsection{The effect of a decreasing velocity on the averaged profiles.}\label{subsec:veloeffect}
We also calculated what happens to the emergent aver-
aged profiles in the strongly dynamic case 
(with $15$ minutes of integration, emulating
an observation) when we gradually reduce the velocity field 
by a constant scaling factor $F$, keeping the rest of atmospheric
magnitudes unperturbed (see Fig. \ref{fig:consistency_test}).
As expected, we find that the polarization amplitudes decrease in
proportion to $F$, from the original case, with $F=1$, towards
the static case, with $F=0$. Note that the core of the
8498 \AA\ line goes through zero for a certain $F$ value (near $F = 0.6$). Thus,
depending on the magnitude
of the velocity gradients, its linear polarization amplitude will be positive or negative. This fact suggests an additional
way to diagnose velocity gradients along the line-of-sight.
However, it is important to keep in mind that this sensitivity
also depends on the variations in density and temperature, as
shown in Sec. \ref{subsec:dynamic}.

Furthermore, the variation of the $Q/I$ amplitudes is not linear with $F$. 
The change is small for small $F$, is larger for intermediate values
of $F$, and again becomes smaller for the
largest $F$, tending to saturation. 

%Then, we see that, even in the presence of more realistic
%physical conditions, the three lines show a behavior 
%that is very similar to that found in Paper I. The small difference is that now the signals lobe in opposition to the enhanced lobe has a lower amplitude than in the FAL-C models (in other words,
%the $\rm{(Q/I)_{pp}}$ is slightly lower). The reason, of course, is that in the present dynamic models,
%the Doppler shifts and the existence of positive and negative parts in the signals 
%produce cancellations when integrating in time.

%the measrument and comparison of this signals in different integration times could tell us ... 

\section{Considerations on the Hanle effect}\label{sec:hanle}
For magnetic field diagnostics with the Hanle effect it is often 
necessary to know the zero-field
polarization reference \citep[e.g., ][]{stenflo94,trujillo_nature04}. That is what we have tried to do in previous sections, calculating and explaining
the temporal evolution of the linear polarization profiles in chromospheric dynamic simulations. Ideally, this
reference has to be computed under the same
thermodynamical and dynamical conditions than in the real Sun but without magnetic field. 
As the Hanle effect often depolarizes the linear polarization signals, the difference between the
observation and the zero-field calculation can be associated to a magnetic field
by adjusting the magnetic field topology and intensity. The key point is that the reference amplitude must be as precise
as possible. If it is imprecise, variations in the Stokes profiles can be
associated to a magnetic field when they were really due
to uncertainties in another magnitudes, like the temperature or the velocity
field. Due to this reason, the fact that the solar chromosphere is a highly dynamic medium 
brings some complications for the use of the Hanle effect as a diagnostic
mechanism. 

A strategy to avoid the above-mentioned problem is
known as the line ratio technique. It consists of finding a pair of spectral
lines whose thermodynamical behaviour is identical but whose sensitivity to the
magnetic field is different in some range of magnetic field intensity or
inclinations \citep[e.g.,][]{stenflo98,manso04}. In that case, the ratio between the polarization amplitudes
should only change due to variations in the magnetic field, thus allowing us to
measure it after a suitable calibration. As shown by \cite{manso10}, 
the main magnetic sensitivity difference among the lines of the Ca II IR triplet is between the ${\lambda}8498$ line
(which is sensitive to field strengths between 0.001 G and 10 G) and any of the ${\lambda}8662$ and ${\lambda}8542$ lines (which react mainly to sub-gauss magnetic fields and up to 10 G in the latter spectral line). Unfortunately, while the line-cores of the 
${\lambda}8662$ and ${\lambda}8542$ lines originate in similar atmospheric layers, 
the ${\lambda}8498$ line-core originates at significantly deeper atmospheric layers (see Figs. 1 and 5).
Nevertheless, we have found useful to plot in Fig. \ref{fig:lineratios} the time evolution  
of the following polarization line ratios:   
\begin{subequations}\label{eq:hanle1}
\begin{align}
\varrho_1 = \frac{(\rm{Q/I})^{8498}_{pp}}{(\rm{Q/I})^{8662}_{pp}}, \label{hanle1a}
\displaybreak[0] \\
\varrho_2 = \frac{(\rm{Q/I})^{8542}_{pp}}{(\rm{Q/I})^{8662}_{pp}},\label{hanle1b}
\end{align}
\end{subequations}
where the super-index indicates the central wavelength of the transition in \AA. 
These quantities were calculated for each simulation considered before
(weakly and strongly dynamic cases). The more stable they are, the more useful they will be
for inferring the magnetic field. 

We obtain that , in average, $\bar{\varrho_1} =0.15\,\pm\, 0.10$ and $\bar{\varrho_1} =0.16\,\pm\, 0.14$
for the strongly and weakly dynamic cases,
respectively (lower panel in Fig. \ref{fig:lineratios}). The sudden
shape variations (including maximum amplitudes passing by zero) of the 8498 \AA\ line induce large 
instantaneous excursions on $\varrho_1$ .
As expected, a more
stable line ratio is obtained for the second pair of transitions, which are precisely the ones that originate at similar chromospheric heights.
We find $\bar{\varrho_2} =1.06\,\pm\, 0.11$ and $\bar{\varrho_2} =1.00\,\pm\, 0.09$ for the strongly and
weakly dynamic cases, respectively (upper panel in Fig. \ref{fig:lineratios}). If we
repeat the calculations setting to zero the micro-turbulent velocity in the
strongly dynamic case we obtain $\bar{\varrho_1} =0.22\,\pm\, 0.09$ and $\bar{\varrho_2}
=0.97\,\pm\, 0.12$ (dashed black lines in Fig. \ref{fig:lineratios}). These results indicate that the $\bar{\varrho_2} $ line ratio shows a relatively stable 
behavior against variations of the velocity and temperature in the solar atmosphere. Consequently, in principle, $\varrho_2$ could be used as a suitable line ratio to estimate the magnetic field from spectropolarimetric observations of the ${\lambda}8662$ and ${\lambda}8542$ lines. 

Regarding the sensitivity of these lines to the magnetic field and their
applicability for the diagnostic of magnetic fields through the Hanle effect, 
several considerations have to be
taken into account.
 First, the micro-turbulent velocity has a small influence on the averaged
amplitudes and line ratios.
 Second, once the magnetic 
field is included in the calculations, the Hanle effect
typically operates at the line center for static cases. However, in a dynamic
situation there is not a preferred line center wavelength. As the maximum of the
absorption and dispersion profiles occurs at different Doppler shifted
wavelengths, the Hanle effect will operate in a small bandwidth around the
line core.
  Third, according to the static calculations by \cite{manso10}, for 
chromospheric magnetic fields stronger than $0.1$ G in the ``quiet'' Sun, the $\rm{Q/I}$ signal of the 8662 \AA\ line is expected to be Hanle
saturated. Thus, variations between $0.1$ and $10$ G could be measured with
$\varrho_2$, being produced by changes in the linear polarization of the
8542 \AA\ line. 
 Unfortunately, the fluctuations we see in Fig. \ref{fig:lineratios} (exclusively due to the dynamics)
have amplitudes of the same order of magnitude than those expected from the investigations of the 
Hanle effect in static model atmospheres (exclusively due to the magnetic field). More realistic results will be obtained when carrying out calculations of the Hanle effect of the Ca {\sc ii} IR triplet in dynamical model atmospheres. In any case, it is clear that for exploiting the polarization of these lines, we need instruments of high polarimetric sensitivity.
%
%The amplitudes of such fluctuations constitutes an intrinsic limitation that depends on the physical situation (atomic and atmospheric models considered). Then, to be more precise, we have to calculate the curves for the Hanle effect that relate the $\rm{Q/I}$ amplitude with the magnetic field intensity and inclination, but including the effect of the macroscopic velocity of the dynamic model under analysis (see a forthcoming publication).
%And fourth and last, the detection of the small differences in the scattering polarization 
%line ratios can only be achieved with instruments of very high polarimetric sensitivity. 

% 

%Interesting changes are
%expected, like for example an incresing sensitivity to the magnetic field strength
%(due to the influence of vertical velocity gradients in the populations) as the
%line of sight approaches disk center.

 %%%%%%%%%%%%%%%%%%%%%%%%%%%%%%%%%%%%%%%%
%%%%%%%%%%%%%%%%%%%%%%%%%%%%%%%%%%%%%%%%
% CONCLUSIONS
%%%%%%%%%%%%%%%%%%%%%%%%%%%%%%%%%%%%%%%%
%%%%%%%%%%%%%%%%%%%%%%%%%%%%%%%%%%%%%%%%
\section{Conclusions}
The results presented in this paper indicate that the vertical velocity gradients caused by the 
shock waves that take place at chromospheric heights in the HD models 
of Carlsson \& Stein (1997; 2002) have a significant influence on the computed scattering polarization
profiles of the Ca {\sc ii} IR triplet. They show changes in the shape of the $Q/I$ profiles of the three IR lines and clear enhancements in their amplitudes, as well as changes in the sign of the $Q/I$ signal of the ${\lambda}8498$ line. Interestingly enough, such modifications with respect to the static case are evident, not only in the temporally resolved $Q/I$ profiles (e.g., see Fig. 2), but also in the temporally-averaged $\langle Q \rangle / \langle I \rangle$ profiles (e.g., see Fig. 4). This is true even with moderate macroscopic plasma velocities, simply due to the presence of strong vertical velocity gradients like the ones produced by shock waves. This may explain why the above-mentioned modifications of the scattering polarization profiles of the Ca {\sc ii} IR triplet are present not only in the strongly dynamic simulation (Carlsson \& Stein 1997) but also in the weakly dynamic one (Carlsson \& Stein 2002).

Our investigation points out that the development of diagnostic methods based on the Hanle effect in the Ca {\sc ii} IR triplet should take into account that the dynamical conditions of the solar chromosphere may have a significant impact on the emergent scattering polarization signals.  
This complication could be alleviated through the application of line ratio techniques. In Sec. \ref{sec:hanle} we have concluded that the ratio between the polarization amplitudes of the ${\lambda}8542$ and ${\lambda}8662$ transitions would be the best line-ratio choice. However, even in the absence of magnetic fields, the small fluctuations we see in the value of such a line ratio in dynamical model atmospheres could be confused with the presence of magnetic fields in the range 
between $0.1$ and $10$ G. Further work is necessary at this point.

In any case, the fact that realistic macroscopic velocity gradients may have a significant impact on the scattering polarization profiles of the Ca {\sc ii} IR triplet is interesting and important for the diagnostic of the solar chromosphere \footnote{The physical mechanism is general, but its observable effects are expected to be significantly less important in broader spectral lines, such as the Ca {\sc ii} K line and Ly$\alpha$.}. On the one hand, it provides a new observable for probing the dynamical conditions of the solar chromosphere (e.g., by confronting observed Stokes profiles with those computed in dynamical models) . On the other hand, the exploration of the magnetism of the quiet solar chromosphere via the Hanle effect in the Ca {\sc ii} IR triplet (either through the forward modeling approach or via foreseeable Stokes inversion approaches) would have to be accomplished without neglecting the possible effect of the atmospheric velocity gradients on the
  atomic level polarization. 

Several points are still unanswered after this work. First, we need to investigate the 
sensitivity to the Hanle effect of the $\rm{Q/I}$ and $U/I$ profiles of the 
Ca {\sc ii} IR triplet using magnetized and dynamical atmospheric models. Second, we have to investigate whether our one-dimensional radiative transfer 
results remain valid after considering 
realistic three-dimensional models, such as those resulting from magneto-hydrodynamical simulations \citep[e.g.][]{sven04,leenaarts09}. We would tentatively expect that
the strong stratification that gravity imposes in the solar atmosphere facilitate shocks
that propagate mainly in the vertical direction, but with a reduced strength, given the increased
degrees of freedom.

Finally, we mention that our results could be of potential interest in other
astrophysical contexts. For instance, the mechanism of polarization enhancement
due to the presence of shocks might well be the explanation for the
changing amplitudes of the linear polarization signals reported in variable pulsating Mira stars \citep{fabas11}.

\acknowledgments 
We are grateful to Rafael Manso Sainz (IAC) for several useful discussions and advice with the radiative transfer computations. 
Financial support by the Spanish Ministry of Economy and Competitiveness through projects AYA2010-18029 (Solar Magnetism and Astrophysical Spectropolarimetry) and CONSOLIDER INGENIO CSD2009-00038 (Molecular Astrophysics: The Herschel and Alma Era) is gratefully acknowledged.

%%%%%%%%%%%%%%%%%%%%%%%%%%%%%%%%%%%%%%%%%%%%%%%%%%%%%%%%%%%%%%%%%
% The bibliography
%%%%%%%%%%%%%%%%%%%%%%%%%%%%%%%%%%%%%%%%%%%%%%%%%%%%%%%%%%%%%%%%%

\bibliographystyle{apj} 
\bibliography{mybib}

\end{document}